\documentclass[12pt]{amsart}
\usepackage{amsfonts,amssymb}
\RequirePackage{ifpdf}
\usepackage{amsmath}
\usepackage{color}
\usepackage{pstcol,pst-node}
\usepackage{graphics}
\usepackage{epic}
\usepackage{curves}

\def\appendix#1{
\addtocounter{section}{1} \setcounter{equation}{0}
\renewcommand{\thesection}{\Alph{section}}
\section*{Appendix \thesection\protect\indent\quad
#1}
}
\renewcommand{\theequation}{\thesection.\arabic{equation}}

\catcode`\@=11
\def\marginnote#1{}

\newcount\hour
\newcount\minute
\newtoks\amorpm
\hour=\time\divide\hour by60 \minute=\time{\multiply\hour by60
\global\advance\minute by-\hour}
\edef\standardtime{{\ifnum\hour<12 \global\amorpm={am}%
        \else\global\amorpm={pm}\advance\hour by-12 \fi
        \ifnum\hour=0 \hour=12 \fi
        \number\hour:\ifnum\minute<10 0\fi\number\minute\the\amorpm}}
\edef\militarytime{\number\hour:\ifnum\minute<100\fi\number\minute}

\def\draftlabel#1{{\@bsphack\if@filesw {\let\thepage\relax
      \xdef\@gtempa{\write\@auxout{\string
          \newlabel{#1}{{\@currentlabel}{\thepage}}}}}\@gtempa \if@nobreak
    \ifvmode\nobreak\fi\fi\fi\@esphack} \gdef\@eqnlabel{#1}}
    \def\@eqnlabel{}
\def\@vacuum{}
\def\draftmarginnote#1{\marginpar{\raggedright\scriptsize\tt#1}}

\def\draft{
  \oddsidemargin -.5truein
  \def\@oddfoot{\footnotesize \sl preliminary draft \hfil
    \rm\thepage\hfil\sl\today\quad\militarytime}
  \let\@evenfoot\@oddfoot \overfullrule 3pt
    \let\label=\draftlabel
    \let\marginnote=\draftmarginnote
  \def\@eqnnum{(\theequation)\rlap{\kern\marginparsep\tt\@eqnlabel}%
    \global\let\@eqnlabel\@vacuum}

  }
\voffset= - 0.5in \hoffset= - 1.0in

\newcommand{\tr}{\,{\rm Tr}\,}

\def\be{\begin{equation}}
\def\ee{\end{equation}}
\def\bea{\begin{eqnarray}}
\def\eea{\end{eqnarray}}
\def\<{\langle}
\def\>{\rangle}

\def\tr{\mathop{\rm{tr}}}

\def\ocomma{{\phantom{\Bigm|}^{\phantom {X}}_{\raise-1.5pt\hbox{,}}\!\!\!\!\!\!\otimes}}

\newtheorem{theorem}{Theorem}[section]
\newtheorem{lm}[theorem]{Lemma}

\theoremstyle{definition}
\newtheorem{example}[theorem]{Example}

\allowdisplaybreaks

\long\def\rem#1{}

\begin{document}

\title[Spectral curves for hypergeometric Hurwitz numbers]
{Spectral curves for  hypergeometric Hurwitz numbers}
\author{Jan Ambj{\o}rn$^\dagger$}\thanks{$^\dagger$Niels Bohr Institute,
Copenhagen University, Denmark, and IMAPP, Radboud University, Nijmengen, The Netherlands.
Email: ambjorn@nbi.dk.}
\author{Leonid O. Chekhov$^{\ast}$}\thanks{$^{\ast}$Steklov Mathematical
Institute and  Interdisciplinary Scientific Center J.-V. Poncelet, Moscow, Russia; Michigan State University, East Lansing, USA. Email: chekhov@mi.ras.ru.}

\begin{abstract}
We consider multi-matrix models that are generating functions for the numbers of branched
covers of the complex projective line ramified over $n$ fixed points $z_i$, $i=1,\dots,n$,
(generalized Grotendieck's dessins d'enfants) of fixed genus, degree, and the ramification
profiles at two points, $z_1$ and $z_n$. 
Ramifications at other $n-2$ points enter the sum
with the length of the profile at $z_2$ and with the total length of profiles at the
remaining $n-3$ points. We find the spectral curve of the model for $n=5$ using the loop equation technique for
the above generating function represented
as a chain of Hermitian matrices with a nearest-neighbor interaction
of the type $\tr M_iM_{i+1}^{-1}$. The obtained spectral curve is algebraic and provides
all necessary ingredients for
the topological recursion procedure producing all-genus terms of the asymptotic expansion of
our model in $1/N^2$. We discuss braid-group symmetries of our model and perspectives of the proposed method.
\end{abstract}

\maketitle

\section{Introduction}\label{s:intro}
\setcounter{equation}{0}

Hurwitz numbers enumerate combinatorial classes of ramified mappings $f:{\mathbb CP^1\to\Sigma_g}$
of the complex projective line onto a Riemann surface of genus $g$. The terms single or double Hurwitz numbers
describe the cases in which ramification profiles (defined by the corresponding Young tableauxes
$\lambda$ or $\lambda$ and $\mu$) are respectively fixed
at one ($\infty$) or two ($\infty$ and $1$) distinct points
whereas we assume the existence of $m$ other distinct ramification points with only simple ramifications.

Generating functions for Hurwitz numbers possess a number of remarkable properties: 
integrability of character expansions was observed in \cite{KMMM} and further developed in~\cite{AlMMN}. 
A.~Yu.~Orlov and Shcherbin \cite{OS}, \cite{Or}  and  Okounkov \cite{Okun2000} and
Pandharipande \cite{OP} identified the exponential of the generating function for
double Hurwitz numbers with a tau-function of the Kadomtsev--Petviashvili (KP) hierarchy. 
Note that the class of models enjoying the KP hierarchy properties happens to be rather wide.
General conditions on KP tau functions were first formulated by Takasaki \cite{Tak} and 
later reformulated by Goulden and Jackson \cite{GJ} using the Plucker relations.

In \cite{AC1}, \cite{AC2}, we considered Hurwitz numbers corresponding to another ramification pattern in which 
we have a fixed number $n$ of ramification points with arbitrarily complex ramification profiles at every point. The case
$n=3$ treated in  \cite{AC1} is in one-to-one correspondence with the celebrated Grothendieck's {\em dessins d'enfants}.
This case was first considered by Orlov and Shcherbin \cite{OS} who demonstrated that
the exponential of the corresponding generating function is a tau function
of the KP hierarchy. Moreover, the authors of \cite{OS} proposed a  description of a wider class of generating functions
for {\em hypergeometric Hurwitz numbers} with $n$ of ramification points in ${\mathbb C}P^1$:
we fix profiles at two of these points, taking the sum over profiles at all other points with weights
proportional to the {\em lengths} of the remaining $n-2$ profiles. Harnad and Orlov \cite{HO}
demonstrated that all these generating functions are in turn tau functions of the KP hierarchy.

The relation of branched coverings of the complex plane to matrix models was established almost immediately upon the progress in both these branches of combinatorics  in the early 90th, and  the first proposal of a matrix-model description of Belyi morphisms was presented by Di Francesco and
Itzykson \cite{DFI}. Their model, inspired by the Kontsevich matrix model, described the two-profile Hurwitz numbers 
for {\em clean} Belyi morphisms (with only square-root branchings at the third point). 
We know now that this model falls outside of the
KP hierarchy class and no reliable method for constructing a large-$N$ asymptotic expansion of this model has been found since its
formulation. On the other hand,  matrix-model descriptions of single Hurwitz numbers were successfully 
developed in a series of papers
by de Mello Koch and Ramgoolam \cite{DMK-1}, \cite{DMK-2} in cases where we restrict the possible orders of branchings; we then obtain the simplest Hermitian one-matrix model with a polynomial potential of degree that is the maximum allowed order of branching. 

The interest in Hurwitz numbers corresponding to Belyi pairs was revived in the mathematical community
by Zograf \cite{Zograf} (see also \cite{KZ}) who provided recursion relations based on a cut-and-joint procedure 
for the generating function of Grothendieck's {\it dessins d'enfants}. 

In \cite{AC1}, we develped a matrix-model description of Belyi morphisms,
clean Belyi morphisms, and two-profile Belyi morphisms: the three
corresponding matrix models
are the standard Hermitian one-matrix model with a logarithmic addition to the potential,
the Kontsevich--Penner matrix model \cite{ChM}, and the generalized Kontsevich model \cite{MMM-1}, \cite{MMM-2}
also with a logarithmic term (it is the BGW model of \cite{MMS}). 
The multi-matrix-model representation for the hypergeometric Hurwitz numbers then followed  \cite{AMMN} 
in the form of a matrix chain, albeit with a complicated interaction between matrices in the chain. We have
proposed another chain-matrix-model description of hypergeometric Hurwitz numbers in \cite{AC2}, also 
with a nonstandard interaction between the matrices in the chain, 
for which we were able to construct the spectral curve (for the case $n=4$).
In our approach, we fix profiles at two ramification points, fix the length of the profile at the third point,
and fix the total length of profiles at other $n-3$ points. In this case, we can
derive the spectral curve equation in the framework of the $1/N^2$-expansion. 
Armed with this spectral curve and two meromorphic 
differentials $dx$ and $\omega_1(x)dx$ on it 
(the second differential comes from the resolvent of the first matrix in the matrix chain)
we can apply the topological recursion procedure for constructing consecutively the whole series of 
the $1/N^2$-expansion, thus obtaining
hypergeometric Hurwitz numbers for complex curves of arbitrary genus $g$ (the corresponding free-energy term is proportional to 
$N^{2-2g}$). 

It turned out that the chain multi-matrix model with the same interaction $\tr M_iM_{i+1}^{-1}$ between neighbour matrices in the chain was proposed
by G. Akemann and collaborators \cite{AIK}, \cite{AKW} in the context of problems of joint probability distribution in quantum informatics. They constructed kernels for such models in terms of Meijer G-functions, which we also briefly mention at the end of Sec.~\ref{s:model}, but did not address problems of constructing spectral curves.

We first recall the famous relation between Belyi pairs and Galois groups.

\begin{theorem}\label{thm:Belyi}{\rm (Belyi, \cite{Belyi})}  A smooth complex algebraic curve $C$ is defined
over the field of algebraic numbers $\overline{\mathbb Q}$ if and only if we have a nonconstant
meromorphic function $f$ defined on $C$ $(f:C\to {\mathbb C}P^1)$ that is ramified only over the points $0,1,\infty\in {\mathbb C}P^1$.
\end{theorem}

A Belyi pair $(C,f)$ is the curve endowed with such a function. Let $g$ be the genus of $C$ and $d$ the degree of $f$. If we take the preimage
$f^{-1}([0,1])\subset C$ of the real line segment $[0,1]\in {\mathbb C}P^1$ we obtain a connected bipartite
 fat graph with $d$ edges. Vertices of this graph are preimages of $0$ and $1$ and the cyclic ordering of edges entering
 a vertex is fixed by orientation of the curve $C$. Grothendieck had then formulated the following lemma.

\begin{lm}\label{lm:Grot} {\rm (Grothendieck, \cite{Grot})}
There is a one-to-one correspondence between the isomorphism classes of Belyi pairs and connected bipartite fat graphs.
\end{lm}

A Grothendieck {\it dessin d'enfant} is correspondingly a connected bipartite fat graph representing a Belyi pair.
It is well known that we can naturally extend the dessin $f^{-1}([0,1])\subset C$ corresponding to a Belyi pair
$(C,f)$ to a bipartite triangulation of the curve $C$. 
%For this, we cut the complex plane along the (real) line
%containing the points $0,1,\infty$ 
%coloring upper half plane white and lower half plane grey. This defines the partition of
%$C$ into white and grey triangles such that white triangles has common edges only with grey triangles. We then consider a dual graph 
%with three types of edges originated from three homotopically nonequivalent paths between the white and the grey triangles (represented by 
%the points $i$ and $-i$ in the complex plane).

\begin{figure}[tb]
%\hspace*{2cm}
%\epsfysize=6cm
%\vskip .2in
{\psset{unit=0.7}
\begin{pspicture}(-3,-2.4)(3,2.4)
\psframe[linecolor=white, fillstyle=solid, fillcolor=yellow](-3.7,0)(3.7,-2.5)
\pcline[linestyle=solid, linewidth=1pt](-3.7,0)(3.7,0)
%\pcline[linestyle=solid, linewidth=2pt](-1,0)(1,0)
\rput(-1,0){\pscircle*{.1}}
\rput(1,0){\pscircle*{.1}}
\psarc[linecolor=white, linestyle=solid, linewidth=10pt](-1,0){2}{60}{300}
\psarc[linecolor=red, linestyle=dashed, linewidth=1.5pt](-1,0){1.8}{60}{300}
\psarc[linecolor=blue, linestyle=solid, linewidth=1pt](-1,0){2.2}{60}{300}
\psarc[linecolor=green, linestyle=dotted, linewidth=2pt](-1,0){1.8}{-60}{60}
\psarc[linecolor=green, linestyle=dotted, linewidth=2pt](-1,0){2.2}{-60}{60}
\psarc[linecolor=white, linestyle=solid, linewidth=10pt](1,0){2}{-120}{120}
\psarc[linecolor=green, linestyle=dotted, linewidth=2pt](1,0){1.8}{-120}{120}
\psarc[linecolor=blue, linestyle=solid, linewidth=1pt](1,0){2.2}{-120}{120}
\psarc[linecolor=green, linestyle=dotted, linewidth=2pt](1,0){1.8}{120}{240}
\psarc[linecolor=red, linestyle=dashed, linewidth=1.5pt](1,0){2.2}{120}{240}
\pcline[linecolor=white, linestyle=solid, linewidth=10pt](0,1.73)(0,-1.73)
\pcline[linecolor=green, linestyle=dotted, linewidth=2pt](0.2,1.73)(0.2,-1.73)
\pcline[linecolor=green, linestyle=dotted, linewidth=2pt](-0.2,1.73)(-0.2,-1.73)
\pscircle[linecolor=white, linestyle=solid, linewidth=6pt](-1,0){2.15}
\pscircle[linecolor=white, linestyle=solid, linewidth=6pt](1,0){2.15}
\pcline[linecolor=white, linestyle=solid, linewidth=6pt](0,1.73)(0,-1.73)
\rput(0,1.73){\pscircle[linecolor=black, fillstyle=solid, fillcolor=white]{.25}}
\rput(0,-1.73){\pscircle*{.25}}
\rput(-3.7,0){\pscircle[linecolor=black, fillstyle=solid, fillcolor=white]{.1}}
\rput(3.7,0){\pscircle[linecolor=black, fillstyle=solid, fillcolor=white]{.1}}
\psframe[linecolor=white, fillstyle=solid, fillcolor=white](-3.9,0.2)(-3.7,-.2)
\psframe[linecolor=white, fillstyle=solid, fillcolor=white](3.9,0.2)(3.7,-.2)
\rput(-1.7,0){\pscircle[linecolor=black, fillstyle=solid, fillcolor=white]{.1}}
\rput(-.5,0){\pscircle[linecolor=black, fillstyle=solid, fillcolor=white]{.1}}
\rput(.5,0){\pscircle[linecolor=black, fillstyle=solid, fillcolor=white]{.1}}
\rput(1.7,0){\pscircle[linecolor=black, fillstyle=solid, fillcolor=white]{.1}}
\rput(-4.3,0){\makebox(0,0)[cc]{\hbox{{$\infty_-$}}}}
\rput(4.3,0){\makebox(0,0)[cc]{\hbox{{$\infty_+$}}}}
%\rput(-1,.4){\makebox(0,0)[cc]{\hbox{{$0$}}}}
%\rput(1,.4){\makebox(0,0)[cc]{\hbox{{$1$}}}}
\rput(1,1.5){\makebox(0,0)[cc]{\hbox{{\small$\Lambda$}}}}
\rput(1,-1.4){\makebox(0,0)[cc]{\hbox{{\small$\overline\Lambda$}}}}
\end{pspicture}
}
\caption{\small The generalized Belyi fat graph $\Gamma_1$ corresponding to
the case of $n=5$ ramification points
($\infty$, $-(1+\sqrt{5})/2$, $0$, $1$, and
$(3+\sqrt{5})/2$ denoted by small white circles)  of Sec.~\ref{s:n5}.
The generalized Belyi pair $(\mathbb CP^1,\hbox{id})$ corresponds to this graph;
$\infty_{\pm}$ indicate directions of approaching the infinite point in $\mathbb CP^1$.
The symbols $\Lambda$ and $\overline \Lambda$ indicate external field insertions in the matrix-model formalism
of Sec.~\ref{s:model}. For example, this graph contributes the term $N^2\gamma_1\gamma_2\gamma_3^2 t_1 \tr(\Lambda\overline\Lambda)$.}
\label{fi:Belyi}
\end{figure}
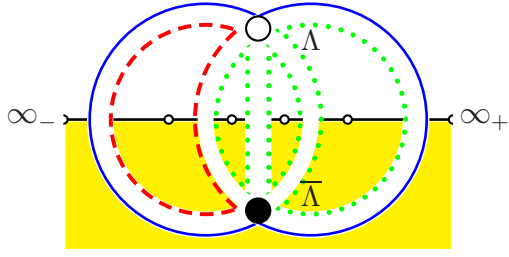

The hypergeometric Hurwitz numbers correspond to
\emph{generalized Belyi pairs}, which are mappings $(f:C\to {\mathbb C}P^1)$ with allowed
ramifications over $n$ fixed points $z_i\in {\mathbb C}P^1$, $i=1,\dots,n$. We then have the splitting of the
curve $C$ into bipartite $n$-gons with edges of $n$ colors (the corresponding fat graphs are then coverings of the
basic graph depicted in Fig.~\ref{fi:Belyi} for $n=5$, which is the main example under consideration in this paper).
We assign color to an edge depending on which of $n$ segments of ${\mathbb R}P^1$---$f^{-1}([\infty_-,z_2])\subset C$,
$f^{-1}([z_2,z_3])\subset C$, $\dots$, $f^{-1}([z_{n-1},z_n])\subset C$,
$f^{-1}([z_n,\infty_+])\subset C$---its image intersects (we identify $z_1$ with the infinity point and let
$\infty_{\pm}$ indicate the directions of approaching this point
along the real axis in $\mathbb CP^1$).
Each face of the dual partition then contains a preimage of exactly
one of the points $z_1,\dots,z_n$, so these faces are of $n$ sorts
(bordered by solid, dotted, or dashed lines in the figure). We call such a graph a {\it generalized Belyi fat graph}.

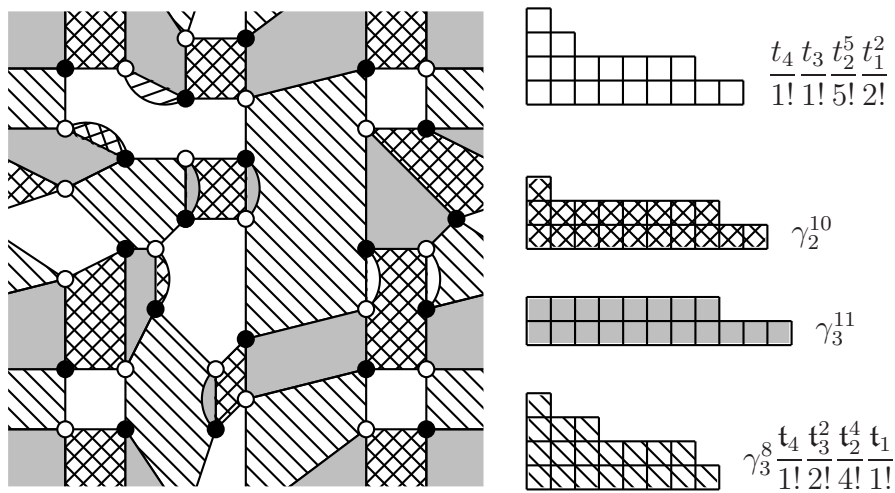
\begin{figure}[tb]
{\psset{unit=0.8}
\begin{pspicture}(-4,-4)(4,5)
\pspolygon[fillstyle=solid,fillcolor=lightgray](-4,-4)(-4,-3)(-3,-3)(-3,-4)(-4,-4)
\rput(7,0){\pspolygon[fillstyle=solid,fillcolor=lightgray](-4,-4)(-4,-3)(-3,-3)(-3,-4)(-4,-4)}
\rput(0,7){\pspolygon[fillstyle=solid,fillcolor=lightgray](-4,-4)(-4,-3)(-3,-3)(-3,-4)(-4,-4)}
\rput(7,7){\pspolygon[fillstyle=solid,fillcolor=lightgray](-4,-4)(-4,-3)(-3,-3)(-3,-4)(-4,-4)}
\pspolygon[fillstyle=vlines](-4,-3)(-4,-2)(-3,-2)(-3,-3)(-4,-3)
\rput(0,5){\pspolygon[fillstyle=vlines](-4,-3)(-4,-2)(-3,-2)(-3,-3)(-4,-3)}
\rput(7,0){\pspolygon[fillstyle=vlines](-4,-3)(-4,-2)(-3,-2)(-3,-3)(-4,-3)}
\rput(7,5){\pspolygon[fillstyle=vlines](-4,-3)(-4,-2)(-3,-2)(-3,-3)(-4,-3)}
\pspolygon[fillstyle=crosshatch](-3,-4)(-3,-3)(-2,-3)(-2,-4)(-3,-4)
\rput(0,7){\pspolygon[fillstyle=crosshatch](-3,-4)(-3,-3)(-2,-3)(-2,-4)(-3,-4)}
\rput(5,0){\pspolygon[fillstyle=crosshatch](-3,-4)(-3,-3)(-2,-3)(-2,-4)(-3,-4)}
\rput(5,7){\pspolygon[fillstyle=crosshatch](-3,-4)(-3,-3)(-2,-3)(-2,-4)(-3,-4)}
\rput(2,4.5){\pspolygon[fillstyle=crosshatch](-3,-4)(-3,-3)(-2,-3)(-2,-4)(-3,-4)}
\rput(2,6.5){\pspolygon[fillstyle=crosshatch](-3,-4)(-3,-3)(-2,-3)(-2,-4)(-3,-4)}
\pspolygon[fillstyle=crosshatch](3,-2)(3,0)(2,0)(2,-2)(3,-2)
\pspolygon[fillstyle=crosshatch](-3,1)(-4.5,0.5)(-6,2)(-5,2)(-3,1)
\rput(8,0){\pspolygon[fillstyle=crosshatch](-3,1)(-4.5,0.5)(-6,2)(-5,2)(-3,1)}
\pspolygon[fillstyle=solid,fillcolor=lightgray](-3,1)(-5,2)(-3,2)(-2,1.5)(-3,1)
\rput(8,0){\pspolygon[fillstyle=solid,fillcolor=lightgray](-3,1)(-5,2)(-3,2)(-2,1.5)(-3,1)}
\rput(8,0){\pspolygon[fillstyle=solid,fillcolor=lightgray](-5,-2)(-3,-2)(-3,-0.5)(-5,-1)(-5,-2)}
\pspolygon[fillstyle=solid,fillcolor=lightgray](-5,-2)(-3,-2)(-3,-0.5)(-5,-1)(-5,-2)
\pspolygon[fillstyle=crosshatch](-3,-2)(-2,-2)(-2,0)(-3,-0.5)(-3,-2)
\pspolygon[fillstyle=vlines](-3,-0.5)(-5,-1)(-5,0)(-4.5,0.5)(-3,-0.5)
\rput(8,0){\pspolygon[fillstyle=vlines](-3,-0.5)(-5,-1)(-5,0)(-4.5,0.5)(-3,-0.5)}
\pspolygon[fillstyle=vlines](-3,1)(-2,0)(-1.5,0)(-1,0.5)(-1,1.5)(-2,1.5)(-3,1)
\pspolygon[fillstyle=vlines](2,3)(2,-1)(0,-1.5)(0,2.5)(2,3)
\pspolygon[fillstyle=solid,fillcolor=lightgray](-2,0)(-1.5,0)(-1.5,-1)(-2,-2)(-2,0)
\pspolygon[fillstyle=solid,fillcolor=lightgray](0,-1.5)(0,-2.5)(2,-2)(2,-1)(0,-1.5)
\pspolygon[fillstyle=vlines](2,-2)(2,-3)(0,-4.5)(0,-2.5)(2,-2)
\rput(0,8){\pspolygon[fillstyle=vlines](2,-3)(0,-4.5)(0,-3)(2,-3)}
\pspolygon[fillstyle=solid,fillcolor=lightgray](2,5)(2,3)(0,2.5)(0,3.5)(2,5)
\rput(0,-8){\pspolygon[fillstyle=solid,fillcolor=lightgray](2,5)(2,3)(0,3.5)(2,5)}
\pspolygon[fillstyle=vlines](-2,-2)(-1.5,-1)(-0.5,-2)(-0.5,-3)(-1,-4.5)(-2,-3)(-2,-2)
\rput(0,8){\pspolygon[fillstyle=vlines](-0.5,-3)(-1,-4.5)(-2,-3)(-0.5,-3)}
\pspolygon[fillstyle=crosshatch](-0.5,-2)(0,-1.5)(0,-2.5)(-0.5,-3)(-0.5,-2)
\pspolygon[fillstyle=solid,fillcolor=lightgray](-2,3)(-1,2.5)(-1,3.5)(-2,5)(-2,3)
\pspolygon[fillstyle=solid,fillcolor=lightgray](3,0)(3.5,0.5)(2,2)(2,0)(3,0)
\rput(0,-8){\pspolygon[fillstyle=solid,fillcolor=lightgray](-2,3)(-1,3.5)(-2,5)(-2,3)}
\rput(-0.47,1){\psarc[fillstyle=solid, fillcolor=lightgray](0,0){0.7}{-45}{45}}
\rput(-1.47,1){\psarc[fillstyle=solid, fillcolor=lightgray](0,0){0.7}{-45}{45}}
\rput(1.53,-0.5){\psarc[fillstyle=solid, fillcolor=white](0,0){0.7}{-45}{45}}
\rput(2.53,-0.5){\psarc[fillstyle=solid, fillcolor=white](0,0){0.7}{-45}{45}}
\rput(-1.97,-0.5){\psarc[fillstyle=crosshatch](0,0){0.7}{-45}{45}}
\rput{65}(-2.65,1.4){\psarc[fillstyle=crosshatch](0,0){0.7}{-60}{60}}
\rput{245}(-1.35,3.1){\psarc[fillstyle=vlines](0,0){0.7}{-60}{60}}
\rput{180}(-0.03,-2.5){\psarc[fillstyle=solid, fillcolor=lightgray](0,0){0.7}{-45}{45}}
\rput(-3,3){\pscircle*(0,0){0.15}}
\rput(-1,2.5){\pscircle*(0,0){0.15}}
\rput(0,3.5){\pscircle*(0,0){0.15}}
\rput(2,3){\pscircle*(0,0){0.15}}
\rput(3,2){\pscircle*(0,0){0.15}}
\rput(2,0){\pscircle*(0,0){0.15}}
\rput(3,-1){\pscircle*(0,0){0.15}}
\rput(2,-2){\pscircle*(0,0){0.15}}
\rput(3,-3){\pscircle*(0,0){0.15}}
\rput(0,-1.5){\pscircle*(0,0){0.15}}
\rput(0,1.5){\pscircle*(0,0){0.15}}
\rput(-0.5,-3){\pscircle*(0,0){0.15}}
\rput(-2,-3){\pscircle*(0,0){0.15}}
\rput(-3,-2){\pscircle*(0,0){0.15}}
\rput(-2,0){\pscircle*(0,0){0.15}}
\rput(-2,1.5){\pscircle*(0,0){0.15}}
\rput(3.5,0.5){\pscircle*(0,0){0.15}}
\rput(-1,0.5){\pscircle*(0,0){0.15}}
\rput(-1,0.5){\pscircle*(0,0){0.15}}
\rput(-1.5,-1){\pscircle*(0,0){0.15}}
\rput(-2,3){\pscircle[fillstyle=solid,fillcolor=white](0,0){0.15}}
\rput(-1,3.5){\pscircle[fillstyle=solid,fillcolor=white](0,0){0.15}}
\rput(0,2.5){\pscircle[fillstyle=solid,fillcolor=white](0,0){0.15}}
\rput(0,0.5){\pscircle[fillstyle=solid,fillcolor=white](0,0){0.15}}
\rput(0,-2.5){\pscircle[fillstyle=solid,fillcolor=white](0,0){0.15}}
\rput(2,2){\pscircle[fillstyle=solid,fillcolor=white](0,0){0.15}}
\rput(3,3){\pscircle[fillstyle=solid,fillcolor=white](0,0){0.15}}
\rput(2,-1){\pscircle[fillstyle=solid,fillcolor=white](0,0){0.15}}
\rput(2,-3){\pscircle[fillstyle=solid,fillcolor=white](0,0){0.15}}
\rput(3,-2){\pscircle[fillstyle=solid,fillcolor=white](0,0){0.15}}
\rput(3,0){\pscircle[fillstyle=solid,fillcolor=white](0,0){0.15}}
\rput(-0.5,-2){\pscircle[fillstyle=solid,fillcolor=white](0,0){0.15}}
\rput(-2,-2){\pscircle[fillstyle=solid,fillcolor=white](0,0){0.15}}
\rput(-3,-3){\pscircle[fillstyle=solid,fillcolor=white](0,0){0.15}}
\rput(-1.5,0){\pscircle[fillstyle=solid,fillcolor=white](0,0){0.15}}
\rput(-1,1.5){\pscircle[fillstyle=solid,fillcolor=white](0,0){0.15}}
\rput(-3,-0.5){\pscircle[fillstyle=solid,fillcolor=white](0,0){0.15}}
\rput(-3,1){\pscircle[fillstyle=solid,fillcolor=white](0,0){0.15}}
\rput(-3,2){\pscircle[fillstyle=solid,fillcolor=white](0,0){0.15}}
\psframe[linecolor=white,fillstyle=solid, fillcolor=white](-6.2,4.5)(-3.95,-4.5)
\psframe[linecolor=white,fillstyle=solid, fillcolor=white](6.2,4.5)(3.95,-4.5)
\psframe[linecolor=white,fillstyle=solid, fillcolor=white](-4.2,3.95)(4.2,5.2)
\psframe[linecolor=white,fillstyle=solid, fillcolor=white](-4.2,-3.95)(4.2,-5.2)
\end{pspicture}
\begin{pspicture}(-2.5,-4)(2.5,4)
%% white tableaux
\pcline(-2,2.4)(1.6,2.4)
\pcline(-2,2.8)(1.6,2.8)
\pcline(-2,3.2)(0.8,3.2)
\pcline(-2,3.6)(-1.2,3.6)
\pcline(-2,4)(-1.6,4)
\pcline(-2,2.4)(-2,4)
\pcline(-1.6,2.4)(-1.6,4)
\pcline(-1.2,2.4)(-1.2,3.6)
\pcline(-0.8,2.4)(-0.8,3.2)
\pcline(-0.4,2.4)(-0.4,3.2)
\pcline(0,2.4)(0,3.2)
\pcline(0.4,2.4)(0.4,3.2)
\pcline(0.8,2.4)(0.8,3.2)
\pcline(1.2,2.4)(1.2,2.8)
\pcline(1.6,2.4)(1.6,2.8)
\rput(2,2.4){\makebox(0,0)[lb]{\hbox{{$\dfrac{t_4}{1!}\dfrac{t_3}{1!}\dfrac{t_2^5}{5!}\dfrac{t_1^2}{2!}$}}}}
%% crosshatched tableaux
\psframe[linecolor=white,fillstyle=crosshatch](-2,0)(2,1.2)
\psframe[linecolor=white,fillstyle=solid, fillcolor=white](-1.6,0.8)(2.1,1.3)
\psframe[linecolor=white,fillstyle=solid, fillcolor=white](1.2,0.4)(2.1,1)
\pcline(-2,0)(2,0)
\pcline(-2,0.4)(2,0.4)
\pcline(-2,0.8)(1.2,0.8)
\pcline(-2,1.2)(-1.6,1.2)
\pcline(-2,0)(-2,1.2)
\pcline(-1.6,0)(-1.6,1.2)
\pcline(-1.2,0)(-1.2,0.8)
\pcline(-0.8,0)(-0.8,0.8)
\pcline(-0.4,0)(-0.4,0.8)
\pcline(0,0)(0,0.8)
\pcline(0.4,0)(0.4,0.8)
\pcline(0.8,0)(0.8,0.8)
\pcline(1.2,0)(1.2,0.8)
\pcline(1.6,0)(1.6,0.4)
\pcline(2,0)(2,0.4)
\rput(2.4,0){\makebox(0,0)[lb]{\hbox{{$\gamma_2^{10}$}}}}
%% gray tableaux
\psframe[linecolor=white,fillstyle=solid, fillcolor=lightgray](-2,-1.6)(2.4,-0.8)
\psframe[linecolor=white,fillstyle=solid, fillcolor=white](1.2,-1.2)(2.5,-0.7)
\pcline(-2,-1.6)(2.4,-1.6)
\pcline(-2,-1.2)(2.4,-1.2)
\pcline(-2,-0.8)(1.2,-0.8)
\multirput(-2,-1.6)(0.4,0){9}{\pcline(0,0)(0,0.8)}
\multirput(1.6,-1.6)(0.4,0){3}{\pcline(0,0)(0,0.4)}
\rput(2.8,-1.6){\makebox(0,0)[lb]{\hbox{{$\gamma_3^{11}$}}}}
%% vlines tableaux
\psframe[linecolor=white,fillstyle=vlines](-2,-4)(1.2,-2.4)
\psframe[linecolor=white,fillstyle=solid, fillcolor=white](-1.6,-2.8)(1.5,-2.3)
\psframe[linecolor=white,fillstyle=solid, fillcolor=white](-0.8,-3.2)(1.5,-2.3)
\psframe[linecolor=white,fillstyle=solid, fillcolor=white](0.8,-3.6)(1.5,-2.3)
\pcline(-2,-4)(1.2,-4)
\pcline(-2,-3.6)(1.2,-3.6)
\pcline(-2,-3.2)(0.8,-3.2)
\pcline(-2,-2.8)(-0.8,-2.8)
\pcline(-2,-2.4)(-1.6,-2.4)
\multirput(-2,-4)(0.4,0){2}{\pcline(0,0)(0,1.6)}
\multirput(-1.2,-4)(0.4,0){2}{\pcline(0,0)(0,1.2)}
\multirput(-0.4,-4)(0.4,0){4}{\pcline(0,0)(0,0.8)}
\pcline(1.2,-4)(1.2,-3.6)
\rput(1.6,-4){\makebox(0,0)[lb]{\hbox{{$\gamma_3^{8}\dfrac{{\mathfrak t}_{4}}{1!}\dfrac{{\mathfrak t}_{3}^2}{2!}
\dfrac{{\mathfrak t}_{2}^4}{4!}\dfrac{{\mathfrak t}_{1}}{1!}$}}}}
\end{pspicture}
}
\caption{\small An example of 19-fold covering of $\mathbb CP^1$ by a torus; we identify the upper and lower as well as left and right boundaries of the square; in the left side we present the corresponding Young tableauxes for collections of polygons of all four colors indicating the weights with which these tableauxes enter the generating function. All separating lines in the right figure are assumed to be double lines.}
\label{fi:19}
\end{figure}

The type of ramification at infinity is fixed by the (unordered) set of solid-line-bounded faces of a generalized
Belyi fat graph: the order of branching is $r$ for a $2r$-gon, so we introduce the generating function that
distinguishes between different types of
branching at infinity, or at $z_1$.
Moreover, we also distinguish between different types of ramifications at the
$n$th point (the point $(3+\sqrt{5})/2$ in Fig.~\ref{fi:Belyi}). We are going therefore to produce 
a \emph{two-profile} generating function for
Hurwitz numbers; branching patterns at two distinct points can be represented by the corresponding Young tableaux.
We let $k_i$ denote the numbers of respective cycles (pre-images of the points $z_i$ on the Riemann surface $C$)
and let $k_1^{(r)}$ and $k_n^{(r)}$ denote the
numbers of cycles of length $2r$ centered at pre-images of the respective points $z_1$ and $z_n$
in a generalized Belyi fat graph. An example of a 19-fold covering of ${\mathbb C}P^1$ by a torus in the case $n=4$ is depicted in Fig.~\ref{fi:19}

As was shown in \cite{AMMN} and \cite{HO}, the exponential of the generating function
\be
{\mathcal F}\bigl[\{t_m\},\{{\mathfrak t}_r\},\gamma_2,\dots,\gamma_{n-1};N\bigr]=\sum_{\Gamma}\frac{1}
{|\hbox{Aut\,}\Gamma|}N^{2-2g}\prod_{r=1}^{\infty}t_{r}^{k_1^{(r)}}
\prod_{s=1}^{\infty}{\mathfrak t}_{s}^{k_n^{(s)}}\prod_{j=2}^{n-1}\gamma_j^{k_j}
\label{gen-fun1}
\ee
is a tau function of the KP hierarchy either in times $t$ or ${\mathfrak t}$. A matrix-model description of this
generating function was proposed in the above papers, and nonlinear cut-and-join equations were constricted in  \cite{AMMN},
but the possibility of solving the model in \emph{topological recursion} terms (see \cite{Ey}, \cite{ChEy}, \cite{CEO})
remained obscure. A matrix model describing a subclass of generating functions (\ref{gen-fun1}) with $\gamma_3=\gamma_4=\cdots=\gamma_{n-1}$ and with arbitrary $\gamma_2>\gamma_3$ was proposed in \cite{AC2}. An advantage of this model is that it can be solved
within a topological recursion method adapted to chain of matrices with nonstandard interaction terms. 
In \cite{AC2}, only the case of one
intermediate matrix ($n=4$) was presented; in the present paper, we extend it to the case $n=5$;  the structure of solution for a 
general $n$ will then become clear. 

We thus solve a matrix model whose free energy is the
generating function
\be
{\mathcal F}\bigl[\{t_m\},\{{\mathfrak t}_r\},\gamma_2,\gamma_{3};N\bigr]=\sum_{\Gamma}\frac{1}
{|\hbox{Aut\,}\Gamma|}N^{2-2g}\prod_{r=1}^{\infty}t_{r}^{k_1^{(r)}}
\prod_{s=1}^{\infty}{\mathfrak t}_{s}^{k_n^{(s)}}\gamma_2^{k_2}\gamma_3^{k_3+\cdots+k_{n-1}},
\label{gen-fun}
\ee
where $N$, $\gamma_2$, $\gamma_3$,
$t_r$, and ${\mathfrak t}_r$
are formal independent parameters and the sum ranges over all (connected) generalized Belyi fat graphs. Our
matrix model  contains an external matrix field
$\Lambda=\hbox{diag\,}(\lambda_1,\dots, \lambda_{\gamma_3N})$ producing the corresponding times
\be
{\mathfrak t}_r=\tr\bigl[(\Lambda\overline\Lambda)^r\bigr].
\label{tt}
\ee

A general case with all $\gamma_j$ different was treated in \cite{AIK}, \cite{AKW}; we leave the problem of finding spectral curves for these generalized models for future studies.

%Sometimes factors $\gamma_1^{k_1}$ and $\gamma_n^{k_n}$ are added but they
%can always be absorbed into the times $t_r$ and ${\mathfrak t}_r$
%by scaling $t_r\to \gamma_1 t_r$ and ${\mathfrak t}_r\to \gamma_n {\mathfrak t}_r$
%for all $r$.

The structure of the paper is as follows.
In Sec.~\ref{s:model}, we show that generating function (\ref{gen-fun}) is the free energy of a special multi-matrix
model represented as a chain of matrices with somewhat nonstandard interaction terms. We describe the
braid-group symmetries of this model. This model can be expressed in terms of an integral over eigenvalues of matrices
from the corresponding matrix chain in a form similar to that of
the standard generalized Kontsevich model (GKM) \cite{MMM-1}, \cite{MMM-2} giving rise to Meijer G-functions.
Our main result is a solution to the loop equations of
this model for $n=5$ in Sec.~\ref{s:n5} describing the spectral curve and two meromorphic differentials on it, i.e., 
all ingredients necessary for constructing corrections in all genera using the topological recursion. For this, we modify
the technique of Eynard and Prats Ferrer \cite{EPF} to evaluate spectral curves for chains of matrices.
We conclude with a discussion of our results.

Throughout the entire text we disregard all multipliers not depending on external fields and times $t_r$; all equalities in the paper must be therefore understood modulo such irrelevant factors.

\section{The matrix model}\label{s:model}
\setcounter{equation}{0}

\subsection{From Hurwitz numbers to the matrix chain}\label{ss:chain}

Our main example in this paper will be hypergeometric Hurwitz numbers for coverings of ${\mathbb C}P^1$ ramified over five points
($n=5$). On the first stage, we take into account the profile at the infinity point for which we contract all solid cycles (centered at
pre-images of $\infty$) assigning the time $t_r$ to every contracted cycle of length $2r$. The thus contracted solid cycles become new vertices. 

The matrix-valued fields $B_i$, $i=2,\dots,n-1$, are general complex-valued matrices such that
$B_2$ is a rectangular matrix of the size $\gamma_2N\times \gamma_3 N$ and we always assume that
$$
\gamma_2>\gamma_3,
$$
and all other matrices $B_3, \dots, B_{n-1}$ are square matrices of the size $\gamma_3N\times \gamma_3 N$.

The matrix-model integral whose free energy is the generating function (\ref{gen-fun}) reads
\be
\int DB_2\cdots DB_{n-1}e^{N\sum_{r=1}^\infty \frac{t_r}{r}\tr
\Bigl[\bigl(
B_2 \cdots B_{n-1}\Lambda \overline \Lambda\,{\overline B}_{n-1}\cdots {\overline B}_2\bigr)^r\Bigr]
-\sum_{j=2}^{n-1}N\tr (B_j\overline B_j)}
\label{model1}
\ee
Here, every vertex contains, besides $2r(n-2)$ insertions of matrix fields we integrate over, $r$ insertions of the external diagonal matrix field
$\Lambda \overline \Lambda$; the factor $1/r$ takes into account the cyclic symmetry of the $2r$-cycle. 

We next perform the variable changing
\be
\begin{array}{l}
{\mathfrak B}_2=B_2B_3\cdots B_{n-1}\cr
{\mathfrak B}_3=B_3\cdots B_{n-1}\cr
\vdots\cr
{\mathfrak B}_{n-1}=B_{n-1}
\end{array}
\label{model2}
\ee
and assume that all matrices ${\mathfrak B}_3,\dots,{\mathfrak B}_{n-1}$
\emph{are invertible} (the matrix ${\mathfrak B}_2$
remains rectangular). With accounting for the Jacobian of transformation (\ref{model2}), the integral (\ref{model1})
becomes
\bea
&{}&\int D{\mathfrak B}_2\cdots D{\mathfrak B}_{n-1}\exp\Bigl\{
-\gamma_2N\tr\log({\mathfrak B}_3{\overline {\mathfrak B}}_3)-\sum_{j=4}^{n-1} \gamma_3N\tr\log({\mathfrak B}_j{\overline {\mathfrak B}}_j)\Bigr.\nonumber\\
&{}&+\sum_{r=1}^\infty N\frac{t_r}{r}\tr\Bigl[({\mathfrak B}_2|\Lambda|^2
{\overline {\mathfrak B}}_2)^r\Bigr]
-N\tr\bigl[{\mathfrak B}_2{\mathfrak B}^{-1}_3{\overline {\mathfrak B}}_3^{-1}
{\overline {\mathfrak B}}_2\bigr]\nonumber\\
&{}&\Bigl.
-N\tr\bigl[{\mathfrak B}_3{\mathfrak B}^{-1}_4{\overline {\mathfrak B}}_4^{-1}
{\overline {\mathfrak B}}_3\bigr]-\cdots
-N\tr\bigl[{\mathfrak B}_{n-2}{\mathfrak B}^{-1}_{n-1}{\overline {\mathfrak B}}_{n-1}^{-1}
{\overline {\mathfrak B}}_{n-2}\bigr]
-N\tr\bigl[{\mathfrak B}_{n-1}{\overline {\mathfrak B}}_{n-1}\bigr]\Bigr\}.
\label{model3}
\eea
This expression justifies restrictions imposed on the matrix sizes: 
we must be able to invert the matrices ${\mathfrak B}_j$ with $j=3,\dots,n-1$
in order to write the corresponding generating function as a free energy of a Hamiltonian of a chain of Hermitian matrices.

We now recall \cite{AKM} that we can write an integral over general complex matrices ${\mathfrak B}_i$
in terms of positive definite Hermitian matrices $X_i$ upon the variable changing
\be
X_i:={\overline {\mathfrak B}}_i {\mathfrak B}_i,\quad i=2,\dots, n-1.
\label{model4}
\ee
All the matrices $X_i$ ($i=2,\dots,n-1$) are of the same size $\gamma_3N\times \gamma_3N$. Changing the integration measure for
rectangular complex matrices introduces a simple logarithmic term (the Marchenko--Pastur law \cite{MP}), and the resulting integral becomes
\bea
&{}&\int {DX_2}_{\ge0}\cdots {DX_{n-1}}_{\ge0}
\exp\Bigl\{ N\sum_{r=1}^\infty \frac{t_r}{r}\tr\bigl[(X_2|\Lambda|^2)^r\bigr]-N\tr (X_2X_3^{-1})
-\cdots-N\tr(X_{n-2}X_{n-1}^{-1})\Bigr.
\nonumber\\
&{}&\quad\quad\Bigl.-N\tr X_{n-1}+(\gamma_2-\gamma_3)N\tr\log X_2-\gamma_2N\tr\log X_3
-\gamma_3N\tr\log (X_4\cdots X_{n-1})\Bigr\}.
\label{model5}
\eea
The logarithmic term in $X_2$ stabilizes the equilibrium distribution of eigenvalues of this matrix in the
domain of positive real numbers; in the case where $\gamma_2=\gamma_3$, we lose this term and must use the technique
of matrix models with hard walls (for a review, see \cite{Ch06}).

Performing a scaling $X_i\to X_i|\Lambda|^{-2}$ for all the integration variables, we bring (\ref{model5}) to a more
familiar form of an integral over a chain of matrices,
\bea
&{}&\int {DX_2}_{\ge0}\cdots {DX_{n-1}}_{\ge0}
\exp\Bigl\{ N\sum_{r=1}^\infty \frac{t_r}{r}\tr(X_2^r)-N\tr (X_2X_3^{-1})
-\cdots-N\tr(X_{n-2}X_{n-1}^{-1})\Bigr.
\nonumber\\
&{}&\quad\quad\Bigl.-N\tr \bigl(X_{n-1}|\Lambda|^{-2}\bigr)+(\gamma_2-\gamma_3)N\tr\log X_2-\gamma_2N\tr\log X_3
-\gamma_3N\tr\log (X_4\cdots X_{n-1})\Bigr\}
\label{model6}
\eea
We use this expression when deriving the spectral curve equation in the next section.

\subsection{The braid-group action}\label{ss:braid}

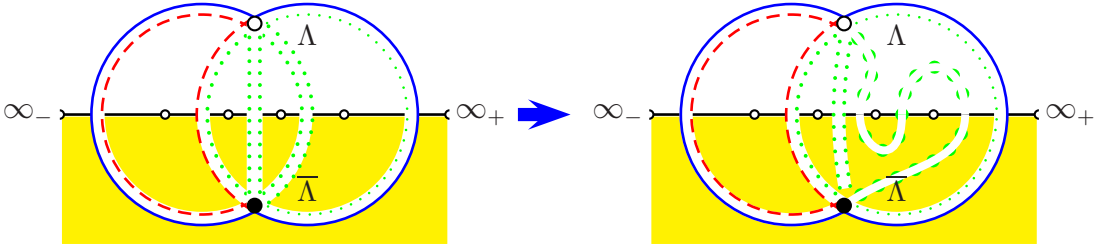
\begin{figure}[tb]
%\hspace*{2cm}
%\epsfysize=6cm
%\vskip .2in
{\psset{unit=0.7}
\begin{pspicture}(-5.5,-2.4)(5.5,2.4)
\psframe[linecolor=white, fillstyle=solid, fillcolor=yellow](-3.7,0)(3.7,-2.5)
\pcline[linestyle=solid, linewidth=1pt](-3.7,0)(3.7,0)
%\pcline[linestyle=solid, linewidth=2pt](-1,0)(1,0)
%\rput(-1,0){\pscircle*{.1}}
%\rput(1,0){\pscircle*{.1}}
%
\pscircle[linecolor=white, linestyle=solid, linewidth=5pt](-1,0){2.1}%{60}{300}
\pscircle[linecolor=white, linestyle=solid, linewidth=5pt](1,0){2.1}%{-120}{120}
\psarc[linecolor=red, linestyle=dashed, linewidth=1pt](-1,0){1.9}{60}{300}
\psarc[linecolor=blue, linestyle=solid, linewidth=1pt](-1,0){2.1}{60}{300}
\psarc[linecolor=green, linestyle=dotted, linewidth=1.5pt](-1,0){1.9}{-60}{60}
\psarc[linecolor=green, linestyle=dotted, linewidth=1.5pt](-1,0){2.1}{-60}{60}
\psarc[linecolor=green, linestyle=dotted, linewidth=1pt](1,0){1.9}{-120}{120}
\psarc[linecolor=blue, linestyle=solid, linewidth=1pt](1,0){2.1}{-120}{120}
\psarc[linecolor=green, linestyle=dotted, linewidth=1.5pt](1,0){1.9}{120}{240}
\psarc[linecolor=red, linestyle=dashed, linewidth=1pt](1,0){2.1}{120}{240}
\pcline[linecolor=white, linestyle=solid, linewidth=6pt](0,1.73)(0,-1.73)
\pcline[linecolor=green, linestyle=dotted, linewidth=1.5pt](0.1,1.73)(0.1,-1.73)
\pcline[linecolor=green, linestyle=dotted, linewidth=1.5pt](-0.1,1.73)(-0.1,-1.73)
%
%\pscircle[linecolor=white, linestyle=solid, linewidth=2pt](-1,0){2.15}
%\pscircle[linecolor=white, linestyle=solid, linewidth=2pt](1,0){2.15}
%\pcline[linecolor=white, linestyle=solid, linewidth=2pt](0,1.73)(0,-1.73)
%
\rput(0,1.73){\pscircle[linecolor=black, fillstyle=solid, fillcolor=white]{.15}}
\rput(0,-1.73){\pscircle*{.15}}
\rput(-3.7,0){\pscircle[linecolor=black, fillstyle=solid, fillcolor=white]{.1}}
\rput(3.7,0){\pscircle[linecolor=black, fillstyle=solid, fillcolor=white]{.1}}
\psframe[linecolor=white, fillstyle=solid, fillcolor=white](-3.9,0.2)(-3.7,-.2)
\psframe[linecolor=white, fillstyle=solid, fillcolor=white](3.9,0.2)(3.7,-.2)
\rput(-1.7,0){\pscircle[linecolor=black, fillstyle=solid, fillcolor=white]{.1}}
\rput(-.5,0){\pscircle[linecolor=black, fillstyle=solid, fillcolor=white]{.1}}
\rput(.5,0){\pscircle[linecolor=black, fillstyle=solid, fillcolor=white]{.1}}
\rput(1.7,0){\pscircle[linecolor=black, fillstyle=solid, fillcolor=white]{.1}}
\rput(-4.3,0){\makebox(0,0)[cc]{\hbox{{$\infty_-$}}}}
\rput(4.3,0){\makebox(0,0)[cc]{\hbox{{$\infty_+$}}}}
%\rput(-1,.4){\makebox(0,0)[cc]{\hbox{{$0$}}}}
%\rput(1,.4){\makebox(0,0)[cc]{\hbox{{$1$}}}}
\rput(1,1.5){\makebox(0,0)[cc]{\hbox{{\small$\Lambda$}}}}
\rput(1,-1.4){\makebox(0,0)[cc]{\hbox{{\small$\overline\Lambda$}}}}
\pcline[linecolor=blue, linewidth=5pt]{->}(5,0)(6,0)
\end{pspicture}
\begin{pspicture}(-5.5,-2.4)(5.5,2.4)
\psframe[linecolor=white, fillstyle=solid, fillcolor=yellow](-3.7,0)(3.7,-2.5)
\pcline[linestyle=solid, linewidth=1pt](-3.7,0)(3.7,0)
%\pcline[linestyle=solid, linewidth=2pt](-1,0)(1,0)
%\rput(-1,0){\pscircle*{.1}}
%\rput(1,0){\pscircle*{.1}}
%
\psarc[linecolor=white, linestyle=solid, linewidth=5pt](-1,0){2}{60}{300}
\pscircle[linecolor=white, linestyle=solid, linewidth=5pt](1,0){2.1}%{-120}{120}
\psarc[linecolor=red, linestyle=dashed, linewidth=1pt](-1,0){1.9}{60}{300}
\psarc[linecolor=blue, linestyle=solid, linewidth=1pt](-1,0){2.1}{60}{300}
%\psarc[linecolor=green, linestyle=dotted, linewidth=1.5pt](-1,0){1.9}{-60}{60}
%\psarc[linecolor=green, linestyle=dotted, linewidth=1.5pt](-1,0){2.1}{-60}{60}
%
\psarc[linecolor=green, linestyle=dotted, linewidth=1pt](1,0){1.9}{-120}{120}
\psarc[linecolor=blue, linestyle=solid, linewidth=1pt](1,0){2.1}{-120}{120}
\psarc[linecolor=green, linestyle=dotted, linewidth=1.5pt](1,0){1.9}{120}{240}
\psarc[linecolor=red, linestyle=dashed, linewidth=1pt](1,0){2.1}{120}{240}
\psarc[linecolor=white, linestyle=solid, linewidth=5pt](8,0){8.1}{170}{190}
\psarc[linecolor=green, linestyle=dotted, linewidth=1.5pt](8,0){8.2}{170}{190}
\psarc[linecolor=green, linestyle=dotted, linewidth=1.5pt](8,0){8}{170}{190}
\psbezier[linewidth=5pt,linestyle=dotted,linecolor=green](0,1.73)(1,0.73)(0.3,0.7)(0.3,0)
\psbezier[linewidth=5pt,linestyle=dotted,linecolor=green](1.1,0)(1.1,-1.1)(0.3,-0.7)(0.3,0)
\psbezier[linewidth=5pt,linestyle=dotted,linecolor=green](1.1,0)(1.1,1.1)(2.3,1.3)(2.3,0)
\psbezier[linewidth=5pt,linestyle=dotted,linecolor=green](0,-1.73)(1,-1)(2.3,-1)(2.3,0)
\psbezier[linewidth=3pt,linecolor=white](0,1.73)(1,0.73)(0.3,0.7)(0.3,0)
\psbezier[linewidth=3pt,linecolor=white](1.1,0)(1.1,-1.1)(0.3,-0.7)(0.3,0)
\psbezier[linewidth=3pt,linecolor=white](1.1,0)(1.1,1.1)(2.3,1.3)(2.3,0)
\psbezier[linewidth=3pt,linecolor=white](0,-1.73)(1,-1)(2.3,-1)(2.3,0)
%
%\pscircle[linecolor=white, linestyle=solid, linewidth=2pt](-1,0){2.15}
%\pscircle[linecolor=white, linestyle=solid, linewidth=2pt](1,0){2.15}
%\pcline[linecolor=white, linestyle=solid, linewidth=2pt](0,1.73)(0,-1.73)
%
\rput(0,1.73){\pscircle[linecolor=black, fillstyle=solid, fillcolor=white]{.15}}
\rput(0,-1.73){\pscircle*{.15}}
\rput(-3.7,0){\pscircle[linecolor=black, fillstyle=solid, fillcolor=white]{.1}}
\rput(3.7,0){\pscircle[linecolor=black, fillstyle=solid, fillcolor=white]{.1}}
\psframe[linecolor=white, fillstyle=solid, fillcolor=white](-3.9,0.2)(-3.7,-.2)
\psframe[linecolor=white, fillstyle=solid, fillcolor=white](3.9,0.2)(3.7,-.2)
\rput(-1.7,0){\pscircle[linecolor=black, fillstyle=solid, fillcolor=white]{.1}}
\rput(-.5,0){\pscircle[linecolor=black, fillstyle=solid, fillcolor=white]{.1}}
\rput(.6,0){\pscircle[linecolor=black, fillstyle=solid, fillcolor=white]{.1}}
\rput(1.7,0){\pscircle[linecolor=black, fillstyle=solid, fillcolor=white]{.1}}
\rput(-4.3,0){\makebox(0,0)[cc]{\hbox{{$\infty_-$}}}}
\rput(4.3,0){\makebox(0,0)[cc]{\hbox{{$\infty_+$}}}}
%\rput(-1,.4){\makebox(0,0)[cc]{\hbox{{$0$}}}}
%\rput(1,.4){\makebox(0,0)[cc]{\hbox{{$1$}}}}
\rput(1,1.5){\makebox(0,0)[cc]{\hbox{{\small$\Lambda$}}}}
\rput(1,-1.4){\makebox(0,0)[cc]{\hbox{{\small$\overline\Lambda$}}}}
\end{pspicture}
}
\caption{\small The  graph transformation resulting in the braid-group transformation permuting the neighbour branching points.}
\label{fi:braid}
\end{figure}

We now address the following natural question: the orders of the ramification points are not fixed \emph{a priori}. So, we must be able to permute matrices in the above matrix chain. This makes our  matrix chain interaction different from the standard matrix chain interaction which is of the form $\tr M_iM_{i+1}$ and which lacks such a symmetry. Indeed, we have a transformation corresponding to an elementary permutation of (neighbour) branching points depicted in Fig.~\ref{fi:braid}. For the matrix chain, this transformation is described by a braid-group generator:
\be
\beta_i:\ \left\{ X_i\to X_{i-1}X_i^{-1}X_{i+1}; \ X_j\to X_j, \ j\ne i \right\}.
\label{betai}
\ee
It is easy to see that the action of each such generator with $3\ge i\ge n-2$ leaves the action (\ref{model6}) invariant.

\subsection{The eigenvalue representation and the KP hierarchy}\label{ss:EVversusKP}

Now we proceed further expressing
integral (\ref{model6}) in terms of eigenvalues $x_i^{(k)}$ of the matrices $X_k$, $k=2,\dots,n-1$.

We apply the Mehta--Itzykson--Zuber integration formula to every term in the chain of matrices in (\ref{model6}).
Taking into account that, for instance, the integral over the unitary group for the term $e^{-N\tr X_kX_{k+1}^{-1}}$
gives
$$
\int DU e^{-N \sum_{i,j=1}^{\gamma_3N}U_{ij}x_i^{(k)}U^*_{ij}[x_j^{(k+1)}]^{-1}}
=\frac{\det_{i,j}[e^{-Nx_i^{(k)}/x^{(k+1)}_j}]}{\Delta(x^{(k)})\Delta(1/x^{(k+1)})}
$$
and that $1/\Delta(1/x^{(k+1)})=\prod_{i=1}^{\gamma_3N}[x_i^{(k+1)}]^{\gamma_3N-1} /\Delta(x^{(k+1)})$
we eventually write the expression in terms of eigenvalues of the matrices $X_k$:
\bea
&{}&\int_0^\infty \prod_{i=1}^{\gamma_3N}dx_i^{(2)}\frac{\Delta(x^{(2)})}{\Delta{\bigl(|\Lambda|^{-2}\bigr)}}
\prod_{k=3}^{n-1}\left(\prod_{i=1}^{\gamma_3N}\frac{dx_i^{(k)}}{x_i^{(k)}}\right)\times\nonumber\\
&{}&\quad\times\prod_{i=1}^{\gamma_3N}\Bigl[\bigl({x_i^{(2)}}/{x_i^{(3)}}\bigr)^{(\gamma_2-\gamma_3)N}
 e^{N\sum_{r=1}^\infty \frac{t_r}{r}(x_i^{(2)})^r-N x_i^{(2)}/x_i^{(3)}-\cdots-Nx_i^{(n-2)}/x_i^{(n-1)}
 -Nx_i^{(n-1)}|\Lambda|_i^{-2}}\Bigr]
 \label{model7}
\eea
Finally, if we introduce logarithmic quantities
$$
\varphi_i^{(r)}=\log x_i^{(r)}, \quad r=3,\dots,n-1,
$$
we can rewrite integral (\ref{model7}) in a more transparent form resembling that of the Toda chain
Lagrangian:
\bea
&{}&\int_0^\infty \prod_{i=1}^{\gamma_3N}dx_i^{(2)}\frac{\Delta(x^{(2)})}{\Delta{\bigl(|\Lambda|^{-2}\bigr)}}
\prod_{i=1}^{\gamma_3N}\biggl[\int_{-\infty}^\infty \prod_{k=3}^{n-1} d\varphi_i^{(k)}\times\biggr.
\nonumber\\
&{}&\quad\times\exp\Bigl[ N\sum_{r=1}^\infty \frac{t_r}{r}\bigl(x_i^{(2)}\bigr)^r
+(\gamma_2-\gamma_3)N\log x_i^{(2)} -(\gamma_2-\gamma_3)N \varphi_i^{(3)}
\Bigr.\nonumber\\
&{}&\quad\quad \biggl.\Bigl.-N x_i^{(2)}e^{-\varphi_i^{(3)}}-Ne^{\varphi_i^{(3)}-\varphi_i^{(4)}}-\cdots
-Ne^{\varphi_i^{(n-2)}-\varphi_i^{(n-1)}}-Ne^{\varphi_i^{(n-1)}}|\Lambda|_i^{-2}\Bigr]\biggr].
\label{model8}
\eea
In this form it is clear that
all integrals w.r.t. $\varphi_i^{(k)}$ are convergent. Performing intermediate integrations w.r.t. variables $\varphi_i^{(k)}$ for fixed $i$ and
$k$ from $3$ to $n-1$ we obtain that every monomial in the Vandermonde determinant in the numerator will be replaced by the corresponding function of the Meijer G-function type, as in \cite{AKW}.

\section{Spectral curve and topological recursion}\label{s:n5}
\setcounter{equation}{0}
In this section, we propose a method for deriving the spectral curve of model (\ref{model6}),
adapting the technique of \cite{EPF} to our case of a nonstandard interaction between matrices in the matrix chain. In this paper,
we restrict ourselves to a technically more transparent case of the three-matrix model given by the integral
\be
\int DM_1\,DM_2\,DM_3\,DM_4\,e^{N\tr[V(M_1)+M_1M_2^{-1}-\gamma_2\log M_2+M_2M_3^{-1}+M_3M_4^{-1}+U(M_4)]},
\label{three-1}
\ee
where the integrations are performed w.r.t.
positive-definite Hermitian matrices of size $\gamma_3N\times \gamma_3N$ and potentials
$V(x)$ and $U(x)$ are two Laurent polynomials of the respective positive degrees $n$ and $r$
(this consideration can be easily generalised to the case where $V'(x)$ and $U'(x)$ are
two rational functions).

The model (\ref{three-1}) satisfies (see \cite{HO}, \cite{HP}) equations of the two-dimensional Toda chain hierarchy,
and in fact these two classes of models are closely related, so solving the
problem of finding the spectral curve in one model can be
translated in a standard way to solving
the corresponding problem in the other model. Because finding spectral curves for multi-matrix models is
more transparent technically than finding spectral curves for models with external matrix fields, 
we stay with the first choice.

\subsection{Field variations and loop equations}\label{ss:loop}

We consider the following variations of the matrix fields $M_i$:
\bea
  \delta M_1&=&\dfrac{1}{x-M_1}\xi([\widehat M_1]),\nonumber \\
  \delta M_2&=&M_2\dfrac{1}{x-M_1}\eta([\widehat M_2]),
  \label{three-2}
\\
  \delta M_3&=&M_3\dfrac{1}{x-M_1}\rho([\widehat M_3]), \nonumber\\
  \delta M_4&=&\dfrac{1}{x-M_1}\chi([\widehat M_4]), \nonumber
\eea
where we take $\xi$, $\eta$, $\rho$, and $\chi$ to be
Laurent polynomials in all but one of arguments $M_i$. We indicate the omitted argument by the symbol $[\widehat M_i]$.
For example, a function $\eta([\widehat M_2])$ is a Laurent polynomial in $M_1$, $M_3$, and $M_4$.
We introduce the standard notation
for the leading term of the $1/N^2$-expansion of the
one-loop mean of the matrix field $M_1$:
\be
\omega_1(x):=\frac 1N\left\langle\tr\frac{1}{x-M_1}\right\rangle_0.
\ee
Here and hereafter, the subscript $0$ of a correlation function indicates
the contribution of the leading order of the $1/N^2$-expansion. A single trace symbol in the brackets
pertains to the whole expression inside the corresponding brackets.

The exact loop equations obtained upon variations (\ref{three-2}) read
\bea
&{}&\frac{1}{N^2}\left\langle\tr\frac{1}{x-M_1}\tr \frac{1}{x-M_1}\xi([\widehat M_1])\right\rangle^{\text{c}}
+\bigl[\omega_1(x)+V'(x)\bigr]\left\langle\tr \frac{1}{x-M_1}\xi([\widehat M_1])\right\rangle\nonumber\\
&{}&\quad +\left\langle\tr\frac{V'(M_1)-V'(x)}{x-M_1}\xi([\widehat M_1])\right\rangle
+\left\langle\tr M_2^{-1}\frac{1}{x-M_1}\xi([\widehat M_1])\right\rangle=0;
\label{i}\\
&{}& \left\langle\tr \frac{-1}{x-M_1}\eta([\widehat M_2])M_2^{-1}M_1\right\rangle
+\left\langle\tr M_3^{-1}M_2\frac{1}{x-M_1}\eta([\widehat M_2])\right\rangle\nonumber\\
&{}&\quad +(\gamma_2-\gamma_3)\left\langle\tr \frac{1}{x-M_1}\eta([\widehat M_2])\right\rangle=0;
\label{ii}\\
&{}& \left\langle\tr \frac{1}{x-M_1}\rho([\widehat M_3])M_3^{-1}M_2\right\rangle
=\left\langle\tr \frac{1}{x-M_1}\rho([\widehat M_3])M_4M_3\right\rangle;
\label{iii}\\
&{}& \left\langle\tr M_3\frac{1}{x-M_1}\chi([\widehat M_4])\right\rangle
+\left\langle\tr U'(M_4)\frac{1}{x-M_1}\chi([\widehat M_4])\right\rangle=0.
\label{iv}
\eea
A complete information on the model is encoded in these loop equations; solving them we can develop the topological
recursion procedure for evaluating all 
terms of the $1/N^2$-expansion. Amazingly enough, the whole topological expansion can be constructed out of the information contained in the leading-order term (the planar approximation, or $g=0$). So, as soon as we derive the spectral curve endowed
with two differentials $dx$ and $\omega_1(x)dx$, we can construct a complete genus expansion in  $N^{2-2g}$ to any desired 
genus $g$ using the \emph{topological recursion} method of \cite{Ey}, \cite{ChEy}, \cite{CEO} (see also \cite{AlMM}; the novel abstract algebraic approach to this method was developed in \cite{ABCO}).
So, finding the spectral curve in terms of $x$ and $\omega_1(x)$ is a definitive step.

Because we obtain the spectral curve in the large-$N$ limit, we disregard the first term in (\ref{i}), which is of
the next order in $1/N^2$. All other terms in all three equations contribute to the leading order.

\subsection{Finding the spectral curve}\label{ss:spectral}

Calculations are rather tedious and consist in substitutions of various functions  $\xi$, $\eta$, $\rho$, and $\chi$ into the above
loop equations. In order to shorten the writing and for the future convenience, we introduce the following notation:
\bea
&{}&{\mathbf a}:=\left\langle\tr \frac{1}{x-M_1}\frac{U'(M_4)-U'(z)}{M_4-z}\right\rangle_0,\nonumber\\
&{}&{\mathbf b}:=\left\langle\tr M_2^{-1}\frac{1}{x-M_1}\frac{U'(M_4)-U'(z)}{M_4-z}\right\rangle_0,\label{abcd}\\
&{}&{\mathbf c}:=\left\langle\tr M_3\frac{1}{x-M_1}\frac{U'(M_4)-U'(z)}{M_4-z}\right\rangle_0,\nonumber\\
&{}&{\mathbf d}:=\left\langle\tr M_2\frac{1}{x-M_1}\frac{U'(M_4)-U'(z)}{M_4-z}\right\rangle_0.\nonumber\\
\eea

We next perform several substitutions enabling us to produce the required identities; in all identities below we
keep only leading terms in the large-$N$ limit.

%\bea
%\rho(M_1,M_2)=M_2^{-1}:&{}&\left\langle\tr \frac{1}{x-M_1}\right\rangle
%+\left\langle\tr M_2^{-1}U'(M_3)\frac{1}{x-M_1}\right\rangle=0,\nonumber\\
%&{}& \omega_1(x)+\left\langle\tr M_2^{-1}U'(M_3)\frac{1}{x-M_1}\right\rangle_0=0;
%\label{three-3}
%\eea
%\bea
%\xi(M_2,M_3)=U'(M_3):&{}&[\omega_1(x)+V'(x)]\left\langle\tr \frac{1}{x-M_1}U'(M_3)\right\rangle_0
%\nonumber\\
%&{}&+\left\langle\tr U'(M_3)\frac{V'(M_1)-V'(x)}{x-M_1}\right\rangle+
%\left\langle\tr M_2^{-1}U'(M_3)\frac{1}{x-M_1}\right\rangle_0=0.\nonumber
%\eea
%Using (\ref{three-3}) and introducing the polynomial
%\be
%P_{n-1}:=\left\langle\tr U'(M_3)\frac{V'(M_1)-V'(x)}{x-M_1}\right\rangle
%\label{Pn-1}
%\ee
%we obtain that
%\be
%\left\langle\tr \frac{1}{x-M_1}U'(M_3))\right\rangle_0=\frac{\omega_1(x)-P_{n-1}(x)}{\omega_1(x)+V'(x)}.
%\ee

{\bf (i)} The first substitution (in (\ref{i})) is
\be
\xi([\widehat M_1])=\frac{U'(M_4)-U'(z)}{M_4-z}:\qquad
 \bigl[\omega_1(x)+V'(x)\bigr]{\mathbf a}+P_{n,m}(x,z)+{\mathbf b}=0,
\label{loop-i}
\ee
where
\be
P_{n,m}(x,z):=\left\langle\tr \frac{U'(M_4)-U'(z)}{M_4-z}\frac{V'(M_1)-V'(x)}{x-M_1}\right\rangle_0
\label{Pnm}
\ee
is a Laurent polynomial in $x$ and $z$ of degrees $n$ and $m$ that are smaller by two than the corresponding degrees of the
potentials $V(x)$ and $U(z)$.

{\bf(ii)} The second substitution (in (\ref{ii})) is
\bea
&{}&\eta([\widehat M_2])=\frac{U'(M_4)-U'(z)}{M_4-z}:\qquad\nonumber\\
&{}& \left\langle\tr \tfrac{-M_1}{x-M_1}\tfrac{U'(M_4)-U'(z)}{M_4-z}M_2^{-1}\right\rangle_0
+\left\langle\tr M_3^{-1}M_2\tfrac{1}{x-M_1}\tfrac{U'(M_4)-U'(z)}{M_4-z}\right\rangle_0
+(\gamma_2-\gamma_3){\mathbf a}=0.
\label{loop-ii}
\eea

{\bf(iii)} The third substitution (in \ref{iii}) is
\bea
&{}&\rho([\widehat M_3])=\frac{U'(M_4)-U'(z)}{M_4-z}:\nonumber\\
 &{}&\left\langle\tr M_3^{-1}M_2\frac{1}{x-M_1}\frac{U'(M_4)-U'(z)}{M_4-z}\right\rangle_0
=\left\langle\tr \frac{1}{x-M_1}\frac{U'(M_4)-U'(z)}{M_4-z}M_4M_3\right\rangle_0.
\label{loop-iii}
\eea

Combining {\bf (ii)} and {\bf (iii)}, we obtain
\be
\widehat Q_m(z) -x{\mathbf b} +(\gamma_2-\gamma_3){\mathbf a}+
\left\langle\tr \frac{1}{x-M_1}\frac{U'(M_4)-U'(z)}{M_4-z}(M_4-z)M_3\right\rangle_0+ z{\mathbf c}=0
\label{inter-1}
\ee
where we have introduced the polynomial
\be
\widehat Q_m(z):=\left\langle\tr \frac{U'(M_4)-U'(z)}{M_4-z}M_2^{-1}\right\rangle_0
\label{hatQ}
\ee
Using the loop equation (\ref{iv}), we can replace $U'(M_4)$ in any expression not containing other insertions of the matrix $M_4$
by $-M_3$, so (\ref{inter-1}) takes the form
\be
-x{\mathbf b} +(\gamma_2-\gamma_3){\mathbf a}+ z{\mathbf c}+\widehat Q_m(z) -
\left\langle\tr \frac{1}{x-M_1}[M_3^2+M_3 U'(z)]\right\rangle_0=0
\label{system-1}
\ee

{\bf(iv)} The fourth substitution (in \ref{i}) is
\bea
&{}&\xi([\widehat M_1])=\frac{U'(M_4)-U'(z)}{M_4-z}M_3:\nonumber\\
 &{}&\bigl[\omega_1(x)+V'(x)\bigr]{\mathbf c}+\widehat P_{n,m}(x,z)+{\mathbf b}
+ \left\langle\tr M_3M_2^{-1}\frac{1}{x-M_1}\frac{U'(M_4)-U'(z)}{M_4-z}\right\rangle_0=0,
\label{loop-iv}
\eea
where
\be
\widehat P_{n,m}(x,z):=\left\langle\tr \frac{U'(M_4)-U'(z)}{M_4-z}\frac{V'(M_1)-V'(x)}{x-M_1}M_3\right\rangle_0.
\label{hatPnm}
\ee

{\bf(v)} The fifth substitution (in (\ref{ii})) is
\bea
&{}&\eta([\widehat M_2])=\frac{U'(M_4)-U'(z)}{M_4-z}M_3:\qquad\nonumber\\
&{}& \left\langle\tr \frac{-M_1}{x-M_1}\frac{U'(M_4)-U'(z)}{M_4-z}M_3M_2^{-1}\right\rangle_0
+{\mathbf d}+(\gamma_2-\gamma_3){\mathbf c}=0.
\label{loop-v}
\eea
Introducing the polynomial in $z$
\be
\widehat{\widehat Q}_m(z):=\left\langle\tr \frac{U'(M_4)-U'(z)}{M_4-z}M_3M_2^{-1}\right\rangle_0
\label{hathatQ}
\ee
we can rewrite (\ref{loop-v}) in the form
$$
\widehat{\widehat Q}_m(z)-x\left\langle\tr \frac{1}{x-M_1}\frac{U'(M_4)-U'(z)}{M_4-z}M_3M_2^{-1}\right\rangle_0
+{\mathbf d}+(\gamma_2-\gamma_3){\mathbf c}=0
$$
and expressing the term in angular brackets using (\ref{loop-iv}), we come to the equation
\be
x\bigl[\omega_1(x)+V'(x)\bigr]{\mathbf c}+(\gamma_2-\gamma_3){\mathbf c}+{\mathbf d}
+x\widehat P_{n,m}(x,z)+\widehat{\widehat Q}_m(z)=0.
\label{system-2}
\ee

The  sixth and the last substitution (in (\ref{i})) is
\be
\xi([\widehat M_1])=\frac{U'(M_4)-U'(z)}{M_4-z}M_2:\qquad
 \bigl[\omega_1(x)+V'(x)\bigr]{\mathbf d}+\widehat{\widehat P}_{n,m}(x,z)+{\mathbf a}=0,
\label{loop-vi}
\ee
where
\be
\widehat{\widehat P}_{n,m}(x,z):=\left\langle\tr \frac{U'(M_4)-U'(z)}{M_4-z}\frac{V'(M_1)-V'(x)}{x-M_1}M_2\right\rangle_0.
\label{hathatPnm}
\ee

We now express ${\mathbf b}$ from (\ref{loop-i}) ending up with the system of three equations (\ref{system-1}), (\ref{system-2}), and
(\ref{loop-vi}) on three variables ${\mathbf a}$, ${\mathbf c}$, and ${\mathbf d}$:
\bea
&{}&\Bigl[x\bigl[\omega_1(x)+V'(x)\bigr]+(\gamma_2-\gamma_3)\Bigr]{\mathbf a}+z{\mathbf c}\nonumber\\
&{}&\qquad\qquad\qquad=-xP_{n,m}(x,z)-\widehat Q_{m}(z)+\left\langle\tr \frac{1}{x-M_1}[M_3^2+M_3 U'(z)]\right\rangle_0\nonumber\\
&{}&\Bigl[x\bigl[\omega_1(x)+V'(x)\bigr]+(\gamma_2-\gamma_3)\Bigr]{\mathbf c}+{\mathbf d}=
-x\widehat P_{n,m}(x,z)-\widehat{\widehat Q}_{m}(z)\label{system-all}\\
&{}&{\mathbf a}+\bigl[\omega_1(x)+V'(x)\bigr]{\mathbf d}=-\widehat{\widehat P}_{n,m}(x,z).\nonumber
\eea

 We now treat the system of equations (\ref{system-all}) as a system of three linear equations on three
 unknowns ${\mathbf a}$, ${\mathbf c}$, and ${\mathbf d}$. We still have a free parameter $z$ and the idea
 is to choose this parameter in the way to make the corresponding system \emph{degenerate}. Then, the condition of
 solvability w.r.t. the right-hand sides of system (\ref{system-all}) produces the equation of the spectral curve.
 
 Introducing the shorthand notation
 \be
 r(x):=x\bigl[\omega_1(x)+V'(x)\bigr]+(\gamma_2-\gamma_3),\qquad s(x):=\omega_1(x)+V'(x)
 \label{rs}
 \ee
 the system determinant is
 $$
 \left| 
\begin{array}{ccc}
r(x)  &  0 &  z \\
 0 &  1 & r(x)  \\
 1 &  s(x) &   0
\end{array}
\right|=-z-r^2(x)s(x),
 $$
 which immediately gives 
 \be
 z=-r^2(x)s(x),
 \label{z}
 \ee
 and the condition of solvability of the degenerate system (\ref{system-all}), or, equivalently, the \emph{equation of the spectral curve} is
 \bea
 &{}&-xP_{n,m}(x,z)-\widehat Q_{m}(z)+\left\langle\tr \frac{1}{x-M_1}[M_3^2+M_3 U'(z)]\right\rangle_0
 \nonumber\\
&{}&\qquad\qquad +r(x) \widehat{\widehat P}_{n,m}(x,z)-s(x)r(x)\bigl[x\widehat P_{n,m}(x,z)+\widehat{\widehat Q}_{m}(z)\bigr]=0,
 \label{spectralcurve}
 \eea
 where $P_{n,m}(x,z)$, $\widehat P_{n,m}(x,z)$, $\widehat{\widehat P}_{n,m}(x,z)$, $\widehat Q_{m}(z)$, and $\widehat{\widehat Q}_{m}(z)$
 are the polynomials given by the corresponding formulas (\ref{Pnm}), (\ref{hatPnm}), (\ref{hathatPnm}), (\ref{hatQ}), and (\ref{hathatQ}),
 $r(x)$ and $s(x)$ are defined in (\ref{rs}), and $z=-r^2(x)s(x)$ (\ref{z}). In order to complete the construction, we need only to evaluate
 the quantities $\left\langle\tr \frac{1}{x-M_1}M_3^k\right\rangle_0$ for $k=1,2$. The method for finding quantities 
 $\left\langle\tr \frac{1}{x-M_1}M_2^k\right\rangle_0$ for any integer $k$ using the loop equation (\ref{i}) alone was presented in
 \cite{AC2}. In the next subsection, we develop this method further and show that we can evaluate similar quantities with insertions of the third, not second, matrix $M_3$ using only two the first loop equations (\ref{i}) and (\ref{ii}). 
 
\subsection{Finding $\left\langle\tr \frac{1}{x-M_1}M_3^k\right\rangle_0$}\label{ss:M3k}

We Introduce the convenient notation
$$
f_{k,r}(x):=\left\langle\tr \frac{1}{x-M_1}M_3^kM_2^r\right\rangle_0, \quad f_{0,0}(x)=\omega_1(x),
$$
and the polynomials and constants
\be
R_{k,r}(x):=\left\langle\tr \frac{V'(x)-V'(M_1)}{x-M_1}M_3^kM_2^r\right\rangle_0,\qquad C_k:=\left\langle\tr M_3^kM_2^{-1}\right\rangle_0.
\label{RC}
\ee
Performing the substitution $\xi([\widehat M_1])=M_3^kM_2^r$ into (\ref{i}), we obtain:
\be
\bigl[\omega_1(x)+V'(x)\bigr] f_{k',r'}(x) +R_{k',r'}(x)+f_{k',r'-1}(x)=0,\quad k',r'\in {\mathbb Z}.
\label{rec-1}
\ee
From the relation (\ref{rec-1}) we can evaluate (see \cite{AC2}) $f_{0,r}(x)$ for any $r\in {\mathbb Z}$ using that $f_{0,0}(x)=\omega_1(x)$.
 
We next consider the substitution $\eta([\widehat M_2])=M_3^k$ in (\ref{ii}). It gives 
$$
\left\langle\tr \tfrac{-M_1}{x-M_1}M_3^kM_2^{-1}\right\rangle_0
+\left\langle\tr \tfrac{-M_1}{x-M_1}M_3^{k-1}M_2\right\rangle_0
+(\gamma_2-\gamma_3)\left\langle\tr \tfrac{-M_1}{x-M_1}M_3^{k}\right\rangle_0=0,
$$
and using the same trick of adding and subtracting $x$ to the nominator of the fraction in the first term, we obtain the second recurrent equation
\be
C_k -x f_{k,-1}(x)+f_{k-1,1}(x)+(\gamma_2-\gamma_3)f_{k,0}(x)=0.
\label{rec-2}
\ee
Using (\ref{rec-1}) and (\ref{rec-2}) we can find $f_{k,0}(x)$, $f_{k,1(x)}$, and $f_{k,-1}(x)$ for all integer $k$. For our purposes here we need only
$f_{k,0}(x)$. Let us fix $k\in {\mathbb Z}$.
We use (\ref{rec-1}) with $(k',r')=(k-1,1)$ to express $f_{k-1,1}(x)$ through $f_{k-1,0}(x)$ and (\ref{rec-1}) with $(k',r')=(k,0)$ to express
$f_{k,-1}(x)$ through $f_{k,0}(x)$. The resulting recursion relation takes the form
\bea
&{}&\Bigl[x\bigl[\omega_1(x)+V'(x)\bigr]+(\gamma_2-\gamma_3)\Bigr]\bigl[\omega_1(x)+V'(x)\bigr]f_{k,0}(x)\nonumber\\
&{}&\qquad\qquad=f_{k-1,0}(x)+R_{k-1,1}(x)+\bigl[\omega_1(x)+V'(x)\bigr](xR_{k,0}(x)+C_k)
\label{rec-3}
\eea
Again, suppling it with the initial condition $f_{0,0}(x)=\omega_1(x)$ we express all $f_{k,0}(x)$ in terms of $r(x)$ and $s(x)$ (\ref{rs})
and polynomials $R_{k',0}(x)$, $R_{k',1}(x)$ and constants $C_{k'}$ (\ref{RC}).

\rem{
\section{MIscellaneous}

It is a standard trick in multi-matrix models to introduce the new variable $y$:
\be
y:=\omega_1(x)+V'(x).
\label{y}
\ee

Despite its complexity even in the simplest cases (say, we obtain a hyperelliptic
curve of maximum genus three for the Gaussian potentials $V(x)$ and $U(z)$ in Example~\ref{ex:1} below),
we still have algebraic curves in all these cases 
in contrast to the case of Hurwitz numbers for branching points with
only simple ramifications for which it was conjectured in \cite{Marino} and shown in \cite{MorSha},
\cite{EyMar} that the corresponding spectral curve
(in the case of one-profile Hurwitz numbers)
is the Lambert curve given by a nonpolynomial equation $x=ye^{-y}$.

\begin{example}\label{ex:1}
Let us consider the case of Gaussian potentials $V(x)=x^2/2$ and $U(z)=z^2/2$. All the polynomials
${P}_{n-1,r-1}$, ${\widehat P}_{n-1,r-1}$, ${\widehat P}_{n-1}$, ${\widehat {\widehat P}}_{n-1}$, and $Q_{r-1}$
are then constants and, moreover, ${P}_{n-1,r-1}=1$ and ${\widehat P}_{n-1,r-1}={\widehat P}_{n-1}$.
Then, after all cancelations, we obtain the spectral curve equation
\be
y-x+{\widehat P}-{\widehat{\widehat P}}y+xy^2+Qy^2+y^2(y-x)(xy+\gamma_2-\gamma_3)=0,
\label{sp}
\ee
which, for the general values of constants in (\ref{sp}), 
describes a hyperelliptic curve of genus three.
\end{example}
}

\section{Conclusion}
We developed further the construction of a matrix  chain representation of the
generating functions for hypergeometric Hurwitz numbers started in \cite{AC2}. Our calculations for the case of $n=5$
distinct branching points demonstrate how we can evaluate the corresponding spectral curve in the case of any fixed $n$; all
these curves are going to be rational. 

There could be several directions of development of this technique. First, and most important one, is that the method proposed in this paper
is still far from being effective: we obtain a rational spectral curve with, albeit finite, but large number of ``free'' parameters. We know that we can actually fix all 
these parameters if we assume the spectral curve to be of genus zero (which is a natural assumption for theories that are perturbative excitations of a Gaussian free-field theory) and if we take into account the asymptotic expansion of the $ydx$ differential at the infinity, which is completely governed by the potential $V'(x)$. 
 
Another direction of development is related to possible applications of  generating functions of type (\ref{gen-fun}) in geometry. It is known that
in the case of so-called clean Belyi morphisms, these functions are related \cite{AC1} to the
free energy of the Kontsevich--Penner matrix model \cite{ChM}, \cite{ChM2}, which is known (\cite{Ch95},\cite{Norbury},\cite{DoNor}) to be the generating function of the numbers of integer points in moduli spaces
${\mathcal M}_{g,s}$ of curves of genus $g$ with $s$ holes with fixed (integer) perimeters. It is tempting to explore possible relations of
these discretization patterns to cut-and-join operators of \cite{Zograf} and \cite{AMMN} in the case of
hypergeometric Hurwitz numbers and to Hodge integrals of \cite{Kazarian}. We can also try to explore a possibility to apply matrix model methods of
our series of papers to studying generalizations of Hurwitz numbers, including weighted Hurwitz numbers \cite{ACEH} and their multispecies generalizations
\cite{Har15}.

\section*{Acknowledgments}
The authors acknowledge support from the ERC Advance Grant 291092 ``Exploring the Quantum Universe'' (EQU)
as well as support of  FNU, the Free Danish Research Council, from the grant ``Quantum Geometry''. 
The work of L.Ch. was supported in part by the Russian Foundation for Basic Research (Grant No. 18-01-00273a).

\def\thetheorem{\Alph{section}.\arabic{theorem}}
\def\theprop{\Alph{section}.\arabic{prop}}
\def\thelemma{\Alph{section}.\arabic{lm}}
\def\thecor{\Alph{section}.\arabic{cor}}
\def\theexam{\Alph{section}.\arabic{exam}}
\def\theremark{\Alph{section}.\arabic{remark}}
\def\theequation{\Alph{section}.\arabic{equation}}

\setcounter{section}{0}

%\appendix{Deriving the Jacobian of transformation (\ref{RR2UVM})}\label{se:notation}
%\setcounter{equation}{0}

\end{document}